\documentstyle[aps]{revtex}
\title{\bf Dynamic and Transport Properties of Dissipative Particle
Dynamics with Energy Conservation.}
\author{J.Bonet Avalos and A.D. Mackie\\
Departament d'Enginyeria
Qu\'{\i}mica, ETSEQ \\
Universitat Rovira i Virgili.\\ Carretera de Salou s/n, 43006
Tarragona (Spain)}

\begin{document}
\maketitle
\parskip 2ex
\renewcommand{\theequation}{\arabic{section}.\arabic{equation}}

\begin{abstract}
Simulation results of the thermal conductivity ${\cal L}$ of 
Dissipative Particle 
Dynamics model with Energy Conservation (DPDE) are reported. We also
present an analysis of the transport equations and the transport
coefficients for 
DPDE based on a local equilibrium approximation. This approach is
valid when 
the particle-particle thermal conductivity $\lambda$ and the friction
coefficient $\zeta$ are large.
A qualitative 
derivation of the scaling form of the kinetic contribution of the
transport of energy is derived, yielding two different forms for the
kinetic contribution to the 
heat transport, depending on the value of $\lambda$.
We find agreement between the theoretically predicted value for ${\cal L}$
and the simulation results, for large $\lambda$ and many
particles interacting at one time. Significant differences are found
for small number of interacting particles, even with large $\lambda$.
For smaller values of $\lambda$,  the obtained macroscopic
thermal conductivity is dominated by diffusive transport, in agreement with
the proposed scaling form.
\end{abstract}

\section{Introduction}

        The model for the simulation of the dynamics of complex systems,
known as {\em Dissipative Particle Dynamics} (DPD), has undergone an
important development in recent years since it was introduced in 1992 by
Hoogerbrugge {\em et al.}\cite{Hoo}. This off-lattice simulation
methodology is especially addressed at the modeling of the behavior of
fluids at a scale where fluctuations are important. Thus, the method is
suitable for both the study of the dynamics of small systems and in
situations dominated by Brownian motion, for instance. The potential of
this methodology was realized from the beginning and, as a consequence,
DPD has already been applied to the study of the dynamics of systems of
practical interest such as polymeric materials\cite{Man1,An1,Man2},
polymer adsorption\cite{Gib}, colloidal suspensions\cite{Lek},
multicomponent systems\cite{Cov}, among others. Despite the advances
carried out
towards a complete understanding of the features of this new methodology,
there are still unclear aspects, mainly concerned with the large-scale
transport properties of the model. The study of these properties, relative
to the extended version of the DPD algorithm which includes heat
transport\cite{ours,ours2}, is the subject of this paper.

        In the DPD model, an ensemble of particles, viewed as mesoscopic
representations of {\em flocks} of physical molecules, interact with each
other through conservative as well as dissipative forces. Random forces
are also explicitly considered to maintain the particles in constant
agitation even in the absence of external force fields. In this method,
the particle-particle interactions are such that the total momentum is
conserved despite the presence of dissipative and random forces and,
therefore, the macroscopic hydrodynamic behavior at long-wavelength and
long-time scales is guaranteed. In addition, since the model is not
defined on a lattice, it is Galilean invariant and has no extra
conservation laws apart from those that are physically relevant. Together
with the 
theoretical advantages, the DPD simulation methodology can be implemented
using only local interactions, which makes it fast to run on a
computer, as in Lattice Gas cellular automata\cite{Lad}, for instance.

        The original DPD method was only concerned with the
conservation 
of the total momentum of the particles so that the model could reproduce
continuity and Navier-Stokes equations at the macroscopic level.
Several refinements, such as to 
demand that the equilibrium momentum distribution of the particles 
be of the
Maxwell-Boltzmann type\cite{Esp1}, were added aiming at a closer relationship
of the model with real physical systems. However, the energy transport and,
therefore, the heat flow, could not be modeled since the DPD model was
essentially {\em isothermal}. 

        An extention of the model
to also incorporate the {\em conservation of the total energy} in the
particle-particle interaction, has been proposed in ref.\cite{ours}, and
later in ref.\cite{Esp2}. More recently, we have refined the
original algorithm to ensure energy conservation at every time-step,
instead of in the mean\cite{ours2}. In our version of the DPD
model with 
energy conservation (referred to as DPDE from now on), together with the
position and momentum, an internal energy
of each particle is explicitly considered as a relevant variable. With the
definition of a particle's internal energy, an energy balance can be
established so that the work done by the dissipative and random forces is
stored or released from the particle's internal energy, and thus the total
energy is held constant in the particle-particle interaction. Therefore,
as a consequence, the addition of the energy conservation to the momentum
conservation in the particle-particle interaction allows the DPDE model
to describe heat transport together with the momentum
transport. Hence, the five hydrodynamic
fields, density, momenta and 
energy, issued from the conservation laws at a microscopic level, can be
properly reproduced by the DPDE model.

        In this paper we address the study of the {\em transport
properties} of the DPDE model from both, theoretical as well as simulation
points of view. In the case of the isothermal DPD, much work has
been devoted to the analysis of the transport
properties of the model, especially with regard to the calculation of the
shear 
viscosity\cite{Ern,Eva,War}, towards a qualitative and quantitative
understanding of the long-wavelength long-time dynamic behavior of the
model. Significant progress has been done in this direction by the formulation
of Boltzmann-type equations, which have been solved by means of
different approximations. In particular, in ref.\cite{Ern} the transport
equations for the macroscopic fields were derived and 
expressions for 
the transport coefficients were given, together with a lucid
discussion of the conceptual implications of the approach. The authors
distinguish a kinetic and a dissipative contribution to the shear
viscosity, essentially proportional to  $1/\gamma$ and $\gamma$,
respectively, with $\gamma$ being the friction coefficient per unit of
mass of the 
particle-particle dissipative force, in that reference. Roughly, the first
one is due to the transport of momentum due to the random motion of the
particles while the second is the contribution due to the direct
frictional force between particles. The qualitative
behavior of the viscosity coefficient was captured by the approach, but
significant deviations between the simulation and the theoretical
predictions were found. The disagreement was found to be more important in 
the region dominated by the kinetic contribution, although theoretical and
simulation results seem to tend to agree in the limit where a given
particle interacts with many others at one time. Such a discrepancy has
been further reported by the simulations of Pagonabarraga et
al.\cite{Pago}. However, very recently, the 
approximations used in the solution of the Boltzmann-like equation of
ref.\cite{Ern} have been reviewed in ref.\cite{War}. Masters {\em et al.}
claim that the discrepancies between 
the theory of ref.\cite{Ern} and the simulations can be explained if the
correlations between the collisions are taken into account. This
is even more dramatic in the case that only a few particles are in
simultaneous interaction with a given one. 

        For the DPDE model, such a deep analysis of the transport
phenomena displayed by the system is still lacking. In ref.\cite{ours2} we
analyzed the equilibrium properties of the model from both the
simulation and the theoretical perspectives. As far as the transport
properties of the model are concerned, in ref.\cite{ours2} we
derived, from simple arguments, expressions for the macroscopic thermal
conductivity and the viscosity coefficients of the DPDE model, in the
limit where the dissipative interactions at the mesoscopic level are
dominant. Furthermore, we have shown figures with convection rolls and
temperature profiles in a DPDE system with cross temperature gradient and
gravity field, proving qualitatively that the model displays the same
features as expected for a fluid.
However, simulation results on these transport coefficients is scarce. Only
data from a model with particles at rest exchanging heat to each other are
available\cite{Esp3}. 

        In the present paper, we do not attempt to derive a complete
theoretical treatment of 
the transport properties of the DPDE model, more complex to handle than
the isothermal DPD model due to, first, a larger number of parameters,
second, the couplings between the different 
macroscopic fields, and third, the dependence of the transport
coefficients with 
respect to the temperature. Instead, we propose a calculation of
the complete set of the transport equations for the
five hydrodynamic fields in the limit in which the {\em local equilibrium}
hypothesis is reasonable. Such a procedure is well known for dissipative
systems\cite{vK,JB4,JB5} and has also been discussed in this context in
ref.\cite{Ern}. This treatment permits us to derive transport
equations for the hydrodynamic fields, and approximate
expressions for the transport
coefficients, valid in the region of parameters where dissipative
contributions are dominant. We report simulation data on the 
thermal conductivity, characteristic of the DPDE model, for different
values of the parameter describing the direct heat transfer between
particles, and for two values of the particle-particle interaction range. 
In all the cases studied, the system can be considered as
{\em non-dilute}, in the 
sense that there are always many particles interacting with each other at
one time. Different behaviors are observed, varying with both, the
magnitude of the dissipative 
coefficient, $\lambda$ in our model, and the {\em interaction number}, $n$
that we define as the number of particles interacting at
one time. We attempt a qualitative explanation of these different
behaviors. Finally, we also report simulation data of the
viscosity of the DPDE model as a 
function of the interaction range. The agreement between the simulation
results and the theoretical predictions for the transport coefficients is
good in the region dominated by 
the dissipative interaction, and for large interaction number, $n$. We
observe, 
however, significant deviations if the interaction number is decreased,
regardless of the value of the dissipative coefficients. The theoretical
explanation of this fact could be related to the recent analysis of
ref.\cite{War} for the isothermal DPD model. In the low $\lambda$ limit, we
observe a particular behavior that we relate to a form of transport due to
particle diffusion.

        The paper is organized as follows. In section 2, we give a brief
review of the algorithm describing the DPDE model used in the simulations.
In section 3, we perform the derivation of the transport equations, based
on a local equilibrium formulation. In this section we give expressions of
the transport coefficients and thermodynamic properties that are compared
with those of ref.\cite{ours2}. In section 4, we show the results obtained
for the thermal conductivity from 
the simulations performed with the DPDE model in different conditions, and
compare with the theoretical predictions. Finally, in section 5 we review
and discuss the main results of this paper.

\section{The DPDE model}

\setcounter{equation}{0}

        The DPDE model is defined from the
stochastic differential equations for the updating of the relevant
variables of the particles\cite{ours,ours2}. This implies equations for
the change in 
positions, $\{\vec{r}_i\}$, and momenta, $\{\vec{p}_i\}$, of all the
particles, according to Newton's 
equations of motion. These interactions include frictional and random
forces due to the
{\em mesoscopic} nature of the particles, issued from a coarse-grained
picture of the physical system represented by the DPDE model. In addition, the
internal energies $\{u_i\}$ of the particles are also considered as
relevant variables. The particle internal energy is necessary in this
description to impose the conservation of the total energy in the
particle-particle interaction, when dissipative and random
forces exist. From a physical point of view, both the random interactions
and the frictional forces, as well as
the internal energy of the particles, arise due to the fact that the DPD
particles may be thought of as if they were subsystems of a large number of
degrees of freedom that are 
not explicitly taken into consideration in the simulation. 

        In ref.\cite{ours2} we have discussed at length the derivation,
the properties and physical consistency of
the algorithms defining the DPDE model. The reader is addressed
to that reference for the details. Thus, let us
consider the dynamics of the DPDE model as that of an ensemble of $N$
particles interacting through conservative,
dissipative and random interactions. Let $\vec{F}_{i}^{ext}$
be the external force acting on the i$^{th}$ particle, and
$\vec{F}_{ij}^C$, the
particle-particle conservative force between the pair $ij$. The frictional
force exerted by the $j^{th}$ particle on the $i^{th}$ one is given by 
\begin{equation}
\vec{F}_{ij}^D = \frac{\zeta_{ij}}{m} \,\hat{r}_{ij} \hat{r}_{ij} \cdot
(\vec{p}_j-\vec{p}_i)   \label{1a}
\end{equation}
which is directed along the vector $\hat{r}_{ij} \equiv
\vec{r}_{ij}/r_{ij}$, with $\vec{r}_{ij}\equiv (\vec{r}_j-\vec{r}_i)$ and
$r_{ij} \equiv |\vec{r}_j-\vec{r}_i|$, 
so that the total angular momentum is also conserved.
$\zeta_{ij}\hat{r}_{ij} \hat{r}_{ij}$ is the 
particle-particle friction tensor, which is a function of the
distance between the particles and also of the particles' temperatures
$\theta_i$, defined later on, through the scalar friction coefficient
$\zeta_{ij}=\zeta_{ij}(r_{ij},\theta_i,\theta_j)$.

        As in the isothermal DPD model, we will also demand that the
particles arrive at a given thermal equilibrium for the momenta, described
by the Maxwell-Boltzmann 
distribution. Thus, together with the dissipative forces, we have to also
introduce random {\em Brownian} forces to compensate, on average, the loss of
energy due to the frictional forces. These random forces are Gaussian and
white, and their amplitudes are obtained from a suitable
fluctuation-dissipation theorem\cite{JB1,ours,ours2}.

        As mentioned, the dissipative particles, as 
mesoscopic views of physical systems, bear a large number of internal
degrees of freedom. We assume that these degrees of freedom relax fast
towards a thermodynamic equilibrium characterized by the particles'
internal energy $u_i$. This permits us to introduce a thermodynamic
description at the particle level and thus define a
particle's temperature, $\theta_i$, as well as a particle's entropy,
$s(u_i)$, related to the internal energy by a {\em particle's equation of
state}
\begin{equation}
\frac{1}{\theta_i} = \frac{\partial s(u_i)}{\partial u_i}       \label{1}
\end{equation}
Dynamically, the $i^{th}$ mesoscopic particle varies its internal energy
due to the work done by frictional and Brownian forces on the other
particles. 
However, we consider that it can also vary its internal energy by
exchanging {\em mesoscopic heat}, $q_{ij}^D$, with 
the $j^{th}$ one, if they have
different particle temperatures:
\begin{equation}
q_{ij}^D = \lambda_{ij} \left(\theta_j-\theta_i \right)
        \label{2} 
\end{equation}
where $\lambda_{ij}= \lambda_{ij}(r_{ij},\theta_i,\theta_j)$ is the
mesoscopic thermal conductivity. Eq. (\ref{2}) recalls the macroscopic
Fourier's Law for heat transport, here defined at a mesoscopic level.
Therefore, the change of the internal energy of a given 
particle is due to both the mesoscopic heat exchange as well as to the
work done by the dissipative forces. In addition, 
we demand that the internal energy be a
fluctuating variable described by classical equilibrium statistical
mechanics\cite{Ca} as that of a small system in contact with a heat bath.
In our 
model, the dynamics of such fluctuations is introduced through a random 
heat flow between the particles, $q_{ij}^R$, that satisfy a given
fluctuation-dissipation theorem\cite{ours,ours2}.

        The algorithm for the DPDE model described so far is obtained
after integration of the stochastic differential equations for the change
in the relevant variables (Langevin-like equations) in a small time step
$\delta t$, and retaining terms up to first order\cite{ours2}. We have
chosen an 
interpretation rule\cite{vK} for these stochastic differential equations
that is neither It\^o nor Stratonovich, but the resulting stochastic
process 
satisfies {\em detailed balance}, a property required for the system to
approach to the proper thermal equilibrium\cite{JB1,Ern,ours2}. One
arrives at an implicit algorithm given by\cite{ours2}
\begin{eqnarray} 
\vec{r}_i^{\, \prime} &=& \vec{r}_i + \frac{\vec{p}_i}{m} \delta t 
                \label{21a} \\
\vec{p}_i^{\, \prime}  &=& \vec{p}_i +\left\{\vec{F}_{i}^{ext} + 
\sum_{j \neq i} \left[\vec{F}_{ij}^C + \frac{\zeta_{ij}}{m} \,
(\vec{p}_j-\vec{p}_i) \cdot \hat{r}_{ij} \hat{r}_{ij} \right] \right\}
\delta t +  
\sum_{j \neq i} \hat{r}_{ij} \Gamma_{ij}^{\prime} \, \delta
t^{1/2} \; 
\Omega_{ij}^{(p)} \label{21b} \\  
u_i^{\prime} &=& u_i + \sum_{j \neq i} \left\{\frac{\zeta_{ij}}{2m^2}
\left[(\vec{p}_j-\vec{p}_i) \cdot \hat{r}_{ij} \right]^2 + \lambda_{ij}
\left(\theta_j-\theta_i \right)
\right\} \delta t  \nonumber \\ 
&+& \sum_{j \neq i} \left\{ \frac{1}{2m} 
(\vec{p}_j^{\, \prime}-\vec{p}_i^{\, \prime}) \cdot 
\hat{r}_{ij} \Gamma_{ij}^{\prime} 
\; \Omega_{ij}^{(p)} + \mbox{Sign}(i-j) \Lambda_{ij}^{\prime}
 \; \Omega_{ij}^{(q)} \right\} \, \delta t^{1/2}      \label{21c}
\end{eqnarray}
where $\delta t$ is the time-step. Here, the prime symbol stands for
the calculation of the respective variables at an advanced instant of time
$t+\delta t$ \cite{vK3}. 
Thus, $\vec{r}_i^{~\prime} \equiv \vec{r}_i
(t+\delta t)$, $\vec{p}_i^{~\prime} 
\equiv \vec{p}_i (t+\delta t)$, and $u_i^{\prime} \equiv u_i (t + \delta t)$
while $\vec{r}_i$, $\vec{p}_i$ and $u_i$ are the value of these functions
at time $t$. $\Gamma_{ij}^{\prime}$ and
$\Lambda_{ij}^{\prime}$ are the amplitudes of the random force and heat
flow given by the corresponding fluctuation-dissipation theorems\cite{ours2}
\begin{eqnarray}
\Gamma_{ij}^2 &=& 2 k \, \Theta_{ij} \, \zeta_{ij} \label{14} \\
\Lambda_{ij}^2 &=& 2 k \, \lambda_{ij} \theta_i \theta_j        \label{14b}
\end{eqnarray}
where we have defined the mean inverse temperature as $\Theta_{ij}^{-1}
=(1/\theta_i+1/\theta_j)/2$. Again, the prime indicates that these
amplitudes need to be calculated with their arguments evaluated at the
advanced time. In eqs. (\ref{21b}) and (\ref{21c}) we
have introduced the random numbers $\Omega_{ij}^{(p)}$ and
$\Omega_{ij}^{(q)}$, which are Gaussian and white, with correlations $\langle
\Omega_{ij}^{(p)} \Omega_{kl}^{(p)} \rangle = \langle
\Omega_{ij}^{(q)} \Omega_{kl}^{(q)} \rangle = (\delta_{ik}\delta_{jl} +
\delta_{il}\delta_{jk})$ and $\langle
\Omega_{ij}^{(p)} \Omega_{kl}^{(q)} \rangle =0 $. These random numbers
arise from the contributions due to the random forces and random heat
flows, respectively. In ref.\cite{ours2} an explicit
algorithm for the DPDE model is also given.

        Although the theoretical predictions are valid for any 
DPDE model, for the simulations done in the present work, however, we have
used the 
model defined by the following functions. For each particle we have used
the equation of state $u_i = \phi \theta_i$, where $\phi$ is a
constant that stands for the heat capacity of a single particle. Notice
that it is convenient to choose $\phi \gg k$ to avoid negative values of the
internal energy of the particle during the simulations.  For the
dissipative interactions we have chosen
\begin{eqnarray}
\zeta_{ij} & = & \zeta_0 \left( 1-\frac{r_{ij}}{r_{\zeta}} \right)^2 \;\;
\mbox{for} \;\; r_{ij} \leq r_{\zeta}; \;\; \mbox{and} \;\; 0 \;\;
\mbox{if} \;\; r_{ij} > r_{\zeta} \label{14c} \\ 
\lambda_{ij} & = & \frac{L_0}{\theta_i \theta_j} \left(
1-\frac{r_{ij}}{r_{\lambda}} \right)^2 \;\; \mbox{for} \;\; r_{ij} \leq
r_{\lambda}; \;\; \mbox{and} \;\; 0 \;\;
\mbox{if} \;\; r_{ij} > r_{\lambda}      \label{14d} 
\end{eqnarray}
where $\zeta_0$ and $L_0$ are constants giving the magnitude of the
mesoscopic friction and thermal conductivity, respectively, and
$r_{\zeta}$ and $r_{\lambda}$ are
the respective ranges of these dissipative interactions. Note that
$\lambda_{ij}$ explicitly depends on the particles' temperatures.
Here, no potential interaction forces are considered.

        Despite the multiple parameters that the algorithm given in eqs.
(\ref{21a}), (\ref{21b}) and (\ref{21c}) displays, we use dimensionless
variables that reduce the relevant parameters to two
\begin{eqnarray}
B &=& \frac{k}{\phi}    \label{14e} \\
C &\equiv& \frac{m L_0}{\zeta_0 \phi T_R^2} \label{14f}
\end{eqnarray}
$B$ measures the relative magnitude of the energy fluctuations with
respect to the average energy of each particle, while $C$ is a ratio
between the characteristic time of momentum fluctuations and energy
fluctuations for one particle. To make the variables dimensionless we have
defined a temperature of reference $T_R$ to be used as
the scale of temperature. Thus, $T = T_R T^*$ and $\theta_i = T_R
\theta_i^*$, where the asterisk is used to denote a dimensionless
variable from now on. The momentum of the particles is made dimensionless
according with $\vec{p}_i =  \vec{p}_i^{~*} \sqrt{2mkT_R}$, using the
characteristic value of the momentum thermal fluctuations
$\sqrt{\langle p_i^2 \rangle}$. The momentum
penetration depth for a pair of particles defines a characteristic length
scale $l =\sqrt{2mkT_R} / \zeta_0$.
Thus, $\vec{r}_i = \vec{r}_i^{~*} l $. The
characteristic scale of time is the relaxation time for the particle's
momentum, that is $t= t^* m/\zeta_0$. The
characteristic scale for the particle's internal energy is chosen to be
$\phi T_R$ so that $u_i = u_i^* \phi T_R$. In summary, when these
dimensionless variables are introduced into the algorithm, the equation
for the momentum and the position remarkably contains no parameters (in
the absence of conservative interactions), and
only depends on the dimensionless range of the frictional force
$r_{\zeta}^*$. In turn, the equation of the energy is
such that the terms related to the mechanical energy
dissipation are proportional to $B$, while the particle-particle heat
transport is proportional to $C$. In addition, the equation is also
functionally dependent on the ranges $r_{\zeta}^*$ and $r_{\lambda}^*$.

\section{Macroscopic properties of the DPDE}

\setcounter{equation}{0}

        The model described so far has a hydrodynamic behavior at
long-wavelengths and long-times, according to the
conservation laws introduced in the particle-particle interaction.
In this section we will derive the macroscopic transport equations of
the model in the local equilibrium approximation, in order to relate the
model parameters with transport coefficients and thermodynamic
relations. Such a procedure yields the dominant contribution of the
transport coefficients in the case in which the dissipative interactions
are large, so that the kinetic transport is subdominant. Under these
conditions, we can expect the system to 
relax fast to the local equilibrium probability distribution function for
the set of variables describing the state of the system, i.e.
\begin{eqnarray}
P_{le}(\{\vec{r}_i\},\{\vec{p}_i\},\{u_i\},t) &\sim&
\exp \left\{-\sum_i
\left(\frac{(\vec{p}_i-m\vec{v}_i)^2}{kT_i}+\frac{\psi^{ext}(\vec{r}_i)}{kT_i}+
\frac{1}{2}\sum_{j\neq i}\frac{\psi(\vec{r}_{ij})}{kT_{ij}}\right)
\right. \nonumber \\
 &+&\left. \sum_i \left(\frac{s_i(u_i)}{k}-\frac{u_i}{kT_i} \right) \right \}
\label{22} 
\end{eqnarray}
We have used the shorthand notation $\vec{v}_i \equiv
\vec{v}(\vec{r}_i,t)$, $T_i=T(\vec{r}_i,t)$ to indicate the macroscopic
velocity and temperature fields, respectively, at the space points
occupied by the i$^{th}$ particle. Furthermore, in the term involving the
particle-particle interaction potential $\psi$ we have written $T_{ij}^{-1}
\equiv (1/T_i+1/T_j)/2$ to maintain the symmetry of the probability
distribution under the exchange $i\rightarrow j$. Corrections to this
probability distribution are essentially
proportional to $1/\zeta$ and to $1/\lambda$, and are neglected in this
approach\cite{JB4}. 

        Let us consider a single component system and ignore external fields
for simplicity. The particle number
density is described by the field in the phase space
\begin{equation}
\rho(\vec{r},t) \equiv \left \langle \sum_{i=1}^N \delta
(\vec{r}-\vec{r}_i) \right \rangle   \label{23}
\end{equation}
where here the average must be taken by means of the instantaneous
non-equilibrium probability distribution, that we will approximate by eq.
(\ref{22}). After differentiating both sides of eq. (\ref{23}) and
averaging, we obtain a
continuity equation reflecting the microscopic particle number
conservation 
\begin{equation}
\frac{\partial}{\partial t} \rho(\vec{r},t) = -\nabla \cdot \left \langle
\sum_{i=1}^N 
\frac{\vec{p}_i}{m} \delta(\vec{r}-\vec{r}_i ) \right \rangle = -\nabla
\cdot \vec{v}(\vec{r},t) \rho (\vec{r},t)       \label{24}
\end{equation}
where, in deriving the first equality of this last equation, use has been
made of eq. (\ref{21a}) written in differential form.

        The momentum density in the system is given by the expression
\begin{equation}
\vec{j}(\vec{r},t) \equiv \left \langle \sum_{i=1}^N \vec{p}_i \delta
(\vec{r}-\vec{r}_i ) \right \rangle \equiv m \rho(\vec{r},t)
\vec{v}(\vec{r},t)  \label{25} 
\end{equation}
where the second equality is in fact a definition of the {\em baricentric}
velocity\cite{dGr} $\vec{v}(\vec{r},t)$. Again,
time differentiating eq.(\ref{25}) and
performing the average with the local-equilibrium 
probability distribution, eq. (\ref{22}), after some algebra one arrives
at  
\begin{eqnarray}
\frac{\partial}{\partial t} m \rho \vec{v} &=& -\nabla \cdot m\vec{v}\vec{v}
\rho  -\nabla kT \rho \nonumber \\
 &+& \int \, d\vec{r}\prime \left[-\Delta \hat{r} \frac{d
\psi}{d \Delta r} + \tilde{\zeta}(\Delta r,\vec{r},\vec{r}\prime) \Delta
\hat{r} \Delta \hat{r} 
\cdot \left(\vec{v}(\vec{r}\prime)-\vec{v}(\vec{r}) \right) \right]
\rho^{(2)}(\vec{r},\vec{r}\prime)  \label{26}
\end{eqnarray}
where we have omitted the explicit time and space-dependence of the macroscopic
fields where no confusion could occur. Here, $\Delta \vec{r} \equiv
\vec{r}\prime-\vec{r}$, $\Delta r \equiv |\Delta 
\vec{r}|$, and $\Delta \hat{r} \equiv \Delta \vec{r} /\Delta r$. In
addition, we have defined the two-particle density from the
relation\cite{Han}
\begin{equation}
\rho^{(2)}(\vec{r},\vec{r\prime},t) \equiv \sum_{i,j\neq i} \left \langle
\delta (\vec{r}-\vec{r}_i ) \delta (\vec{r}\prime-\vec{r}_j ) \right
\rangle \equiv  \rho(\vec{r},t)\rho(\vec{r} \prime ,t) g(\Delta
r,\vec{r},\vec{r}\prime )
        \label{27} 
\end{equation}
where the second equality is, in fact, a definition of the pair distribution
function $g(\Delta r,\vec{r},\vec{r}\prime )$. For this
function we have explicitly written its dependence with respect to $\Delta
\vec{r}$ due to the interparticle potential, together with the $\vec{r}$
and $\vec{r}\prime$-dependence due to the possible spatial variation of
the temperature 
field occuring in eq. (\ref{22}). It is important to realize that
$g(\Delta r, \vec{r}, \vec{r} \prime )$ is invariant under the exchange
$\vec{r} \rightarrow \vec{r} \prime $. Furthermore, in eq. (\ref{26}) we
have introduced the energy-averaged friction coefficient, resulting from a
possible temperature dependence in the mesoscopic friction $\zeta$,
according to 
\begin{equation}
\tilde{\zeta}(\Delta r,\vec{r},\vec{r}\prime) \equiv \frac{1}{A^2}\int \,
du_i du_j \; \zeta_{ij}(\Delta r;\theta_i(u_i),\theta_j(u_j)) \,
e^{s(u_i)/k-u_i/kT_i}e^{s(u_j)/k-u_j/kT_j} 
                \label{27b}
\end{equation}
where $A$ is a normalization factor. Since all the pairs of particles
are equivalent, we have dropped the subscript $ij$ on the left hand side of
this last equation. 
Again, the explicit $\vec{r}$ and
$\vec{r}\prime$-dependences come from the spatial variation of the macroscopic
temperature field and is also invariant under the exchange $\vec{r}
\rightarrow \vec{r} \prime $ by construction of $\zeta$. To obtain the
long-wavelength behavior of eq.
(\ref{26}), we use the fact that the integrand in eq. (\ref{26}) is
significantly different from zero only
in a small neighborhood of $\vec{r}\prime$ around
$\vec{r}\prime = \vec{r}$, because $\tilde{\zeta}$
is a function of short range with respect to $\Delta r$. We thus
expand the $\vec{r} \prime$ dependence of the 
velocity, temperature and the density fields in powers of $\Delta
\vec{r}$, up to the first significant order. Then, after some algebra, eq.
(\ref{26}) can be cast under the form
\begin{equation}
\frac{\partial}{\partial t} m \rho \vec{v} = -\nabla \cdot m\vec{v}\vec{v}
\rho  -\nabla p + \nabla \cdot \vec{\vec{\sigma}}      \label{28}
\end{equation}
which is that of a Navier-Stokes equation. Here, we have defined the {\em
pressure} exerted by the DPDE
system
\begin{equation}
p(\vec{r},t) \equiv kT \rho(\vec{r},t) - \frac{2 \pi}{3} \rho^2
\int_0^{\infty} d\Delta r \; \Delta r^3 \, \frac{\psi(\Delta r)}{\partial
\Delta r}  g(\Delta r, \vec{r},\vec{r})
                        \label{29}
\end{equation}
Note that in this expression the last argument of $g(\Delta r,
\vec{r},\vec{r})$ is $\vec{r}$ and not $\vec{r} \prime$. Eq. (\ref{29}) is
formally identical to that obtained from 
thermodynamic considerations for a physical system with
pair interaction potentials\cite{Han}. A similar expression has been
also obtained in ref.\cite{Ern} for the isothermal DPD model, and is 
given in eq.(3.10) of ref.\cite{ours2} for the DPDE model. We have also
defined the macroscopic stress tensor field
\begin{equation}
\vec{\vec{\sigma}} \equiv \left(2 \eta_4 \rho^2
\stackrel{\circ}{\overline{\nabla \vec{v}}}+\frac{5}{3}\eta_4 \rho^2
\vec{\vec{1}} \nabla \cdot \vec{v} \right)      \label{29b}
\end{equation}
From the coefficient of the symmetric and traceless velocity gradient,
$\stackrel{\circ}{\overline{\nabla \vec{v}}}$, we identify the shear viscosity
coefficient
\begin{equation}
\eta = \rho^2 \eta_4  \; \; \mbox{with} \; \; \eta_4 \equiv  \frac{2
\pi}{15} \int_0^{\infty} 
d\Delta r \; \Delta r^4 \,  \tilde{\zeta} (\Delta r,\vec{r},\vec{r})
g(\Delta r,\vec{r},\vec{r})         \label{30}
\end{equation}
The coefficient $\eta_4$ is a function of the position and
of the time 
through the temperature field. The volume viscosity is
obtained from the coefficient of the $\nabla \cdot 
\vec{v}$-term, and reads
\begin{equation}
\eta_v = \frac{5}{3} \eta_4 \, \rho^2   \label{31}
\end{equation}
Both viscosity coefficients agree with the dominant contribution of the
transport 
coefficients given in ref.\cite{Ern}, in the limit of negligible kinetic
momentum transport. Here, however, these viscosity coefficients can be
functions of the temperature through $\tilde{\zeta}$ due to the possible
dependence of $\zeta$ on the particles' temperatures. Note
that neither kinetic nor particle-particle interaction
potential contributions are significant to the momentum transport
coefficients in the large friction limit developed here.

        The calculation of the energy transport equation requires the
calculation of the macroscopic equations for the 
transport of each of the three
contributions to the total energy density, that is, the kinetic, 
potential, and internal energy densities. 
Let us first analyze the change in
the mechanical energy density, given by the sum of the first two mentioned
contributions. The calculation follows the
same lines as in the case of the momentum transport, that is, 
we first time-differentiate the proper variable, 
\begin{equation}
\rho (\vec{r},t) e_m(\vec{r},t)
\equiv  \sum_i \left \langle
\left [\frac{p_i^2}{2m} + \frac{1}{2} \sum_{j\neq i}
\psi(\vec{r}_{ij})\right] \delta 
(\vec{r}-\vec{r}_i ) \right \rangle    \label{32a}
\end{equation}
in this case, and then expand the result in the gradients of the
macroscopic fields. One finally arrives at
\begin{equation}
\frac{\partial}{\partial t} \rho e_m = -\nabla \cdot \left( \frac{1}{2} m \rho
\vec{v}^2 + \rho \Psi + \frac{3}{2} \rho kT \right) \vec{v} -\nabla \cdot
\vec{v} p +\vec{v} \cdot \left ( \nabla \cdot \vec{\vec{\sigma}} \right)
       \label{32}
\end{equation}
where 
\begin{equation}
\Psi (\vec{r},t) \equiv 2 \pi \rho \int d \Delta r \, \Delta r ^2 g(\Delta r)
\psi (\Delta r)  \label{33}
\end{equation}
is the macroscopic potential energy density. The behavior described by
eq. (\ref{32}) was already found in ref.\cite{Ern} for the long-time
behavior of the mechanical energy of the isothermal DPD model.

        The equation describing the transport of the particle's internal
energy is found by analyzing the evolution of the density 
\begin{equation}
\rho e_u \equiv
\left \langle \sum_i u_i \delta
(\vec{r}-\vec{r}_i ) \right \rangle \label{35aa}
\end{equation}
Time-differentiating this variable and
averaging, one finds
\begin{eqnarray}
\frac{\partial}{\partial t} \rho e_u = -\nabla \cdot \rho e_u \vec{v} &+&
\int d\vec{r} \prime \left \langle \left\{ \frac{1}{2m^2}
\left[\left(\vec{p}_j-\vec{p}_i \right) \cdot \Delta \vec{r} \right]^2
\zeta (\Delta r, \theta_i,\theta_j)
+ \lambda (\Delta r,
\theta_i,\theta_j) (\theta_j-\theta_i) \right\} \right. \nonumber \\
 & & \left.  \sum_{i,j\neq i} 
\delta (\vec{r}-\vec{r}_i ) \, \delta (\vec{r} \prime
-\vec{r}_j )\right \rangle    \label{35a}
\end{eqnarray}
As before, the integrand is only different from zero for small values of
$\Delta \vec{r}$, which allows us to expand all of the $\vec{r}
\prime$-dependence of the integrand around $\vec{r}$. Up to the first
significant order we obtain the transport equation for this form of energy
\begin{equation}
\frac{\partial}{\partial t} \rho e_u = -\nabla \cdot \rho e_u \vec{v} +
\vec{\vec{\sigma}} : \nabla \vec{v} +\nabla
\cdot {\cal L} \nabla T    \label{35} 
\end{equation}
The first term on the right hand side of eq. (\ref{35}) is the particle
internal energy advected by the mean flow, the second term is a source of
particle internal energy due to the so-called viscous heating due to the
mechanical energy dissipated by viscous forces. The last term in this
equation is the macroscopic energy transport due to heat 
flow. Let us analyze this heat flow term in more depth. First of all, we
define the  
energy-averaged function $\tilde{\lambda}$, according to the relation
\begin{equation}
\tilde{\lambda}(\Delta r,\vec{r},\vec{r} \prime) (T(\vec{r}
\prime)-T(\vec{r}) ) 
\equiv \frac{1}{A^2}\int \, du_i du_j \; 
\lambda(\Delta r;\theta_i,\theta_j) (\theta_j-\theta_i) \,
e^{s(u_i)/k-u_i/kT_i}e^{s(u_j)/k-u_j/kT_j}      \label{35b}
\end{equation}
as in eq. (\ref{27b}). Thus, from eq. (\ref{35a}) and eq. (\ref{35b}) we
can write 
\begin{eqnarray}
\lefteqn{\int d\vec{r} \prime \left \langle \lambda (\Delta r, \theta_i,\theta_j) 
(\theta_j-\theta_i)  \sum_{i,j\neq i} 
\delta (\vec{r}-\vec{r}_i ) \, \delta (\vec{r} \prime
-\vec{r}_j )\right \rangle =} \nonumber \\
 & & \int d\vec{r} \prime \, \tilde{\lambda}(\Delta
r,\vec{r},\vec{r} \prime) [T(\vec{r} \prime)-T(\vec{r}) ]
\rho^{(2)}(\vec{r},\vec{r} \prime)          \label{35c}
\end{eqnarray}
Expanding now the $\vec{r} \prime $-dependence around $\vec{r}$ in powers
of $\Delta \vec{r}$ and
collecting the first significant order, the right hand side of eq.
(\ref{35c}) becomes
\begin{equation}
\nabla \cdot \rho ^2 \left\{\frac{2 \pi}{3} \int_0^{\infty} d\Delta r \,
\Delta r ^4 \, \tilde{\lambda}(\Delta 
r,\vec{r},\vec{r}) g(\Delta r, \vec{r}, \vec{r}) \right\} \nabla T
                \label{35d} 
\end{equation}
Thus, the macroscopic thermal conductivity takes the form 
\begin{equation}
{\cal L} \equiv \rho ^2 \left\{\frac{2 \pi}{3} \int_0^{\infty}
d\Delta r \, \Delta r ^4 \, \tilde{\lambda}(\Delta 
r,\vec{r},\vec{r}) g(\Delta r, \vec{r}, \vec{r}) \right\} \equiv \rho^2
\lambda_4       \label{35e}
\end{equation}
where the coefficient $\lambda_4$ is analogous to $\eta_4$ defined in eq.
(\ref{30}). This result has been given in eq.(3.25) of ref.\cite{ours2}.

        The total energy transport equation is obtained by adding eqs.
(\ref{32}) and (\ref{35}), yielding
\begin{equation}
\frac{\partial }{\partial t} \rho e = -\nabla \cdot \left(\frac{1}{2}
m\rho \vec{v} + \rho \varepsilon  \right) - \nabla \cdot p \vec{v} + \nabla
\cdot \left(\vec{v} \cdot \vec{\vec{\sigma}} \right) + \nabla \cdot
{\cal L} \nabla T        \label{36}
\end{equation}
where $e = e_m+e_u$ and we have introduced the {\em macroscopic} internal
energy density of the DPDE particles 
system, $\varepsilon$, as the sum of the internal kinetic energy, the
interaction 
potential energy between the particles and the internal energy of the
particles 
\begin{equation}
\rho \varepsilon \equiv \frac{3}{2} kT \rho + \rho \Psi + \rho e_u
        \label{37} 
\end{equation}
which is the same equation of state as given in eq. (3.12) of
ref.\cite{ours2}. 
Note that Eq. (\ref{36}) corresponds to a transport equation for a
conserved magnitude. This form reflects the fact that the 
total energy of the system is conserved in the particle-particle
interaction. Finally, making use of eq. (\ref{28}) to eliminate the
macroscopic kinetic energy transport, one arrives at the transport
equation for the macroscopic internal energy 
\begin{equation}
\frac{\partial}{\partial t} \rho \varepsilon = -\nabla \cdot \rho
\varepsilon \vec{v} - p \nabla \cdot \vec{v} + \vec{\vec{\sigma}} : \nabla
\vec{v} + \nabla \cdot {\cal L} \nabla T         \label{38}
\end{equation}
Each term on the right hand side of this equation has an interpretation in
the framework of the macroscopic transport phenomena. The first term
stands for the advection of the macroscopic internal energy due to
macroscopic fluid motion. The second and third terms are, respectively, a
change in the internal energy due to the work due to an expansion of the
fluid, and the viscous heating of the system due to viscous forces. From
the fourth and last term we can define the {\em macroscopic heat
flow}
\begin{equation}
\vec{J}_q \equiv -{\cal L} \nabla T       \label{39}
\end{equation}
which is precisely the macroscopic expression of the Fourier law.

        Although the theoretical approach developed here is
only applicable to the large friction limit, we end this section
by including a qualitative evaluation of the kinetic contributions to the
thermal conductivity, to enable a comparison
with the simulation data. 

        Effectively, we can estimate the time $t_p$ taken by a particle to
loose the memory of its initial momentum as being given by 
\begin{equation}
t_p \sim \frac{m}{\zeta_0 \rho r_{\zeta}^3}   \label{45}
\end{equation}
as follows from a balance between $dp_i/dt \sim p_i/t_p$
and the 
force given in eq. (\ref{1a}), times the number of particles inside the
interaction range, $\rho r_{\zeta}^3$. Since the particles have an average
velocity $\overline{v} \sim \sqrt{kT/m}$, we can estimate the momentum
penetration depth as
\begin{equation}
l_p \sim \overline{v} t_p \sim \frac{\sqrt{mkT}}{\zeta_0 \rho r_{\zeta}^3}
        \label{45b} 
\end{equation}
On the other hand, we can also estimate the characteristic relaxation time
for the internal energy of a given particle (much larger than the kinetic
energy since we assume that $\phi \gg k$) from a  
balance between $du_i/dt \sim \phi
T/t_q$ and the mesoscopic heat flow, according to eq. (\ref{2})
This leads to
\begin{equation}
t_q \sim \frac{\phi T^2}{L_0 \rho r_{\lambda}^3}     \label{45c}
\end{equation}
where we have also considered that $\rho r_{\lambda}^3$ particles are
simulatenously interacting.
Hence, if $t_q \ll t_p$ the particle covers a distance $l_q$ given by
\begin{equation}
l_q \sim \overline{v} t_q \sim \sqrt{\frac{kT}{m}}\frac{\phi T^2}{L_0
\rho r_{\lambda}^3}     \label{45d}
\end{equation}
in the time the energy is dissipated. Thus, if $l_q \ll l_p$, the energy
transport is dominated by the dissipative contribution and the
energy is delivered before the momentum has relaxed. Thus, the
contribution to the energy flow due to the
motion of the particles is given by\cite{Lif}
\begin{equation}
J_q \sim \phi \rho \overline{v} l_q \nabla T         \label{45e}
\end{equation}
Therefore, in the region dominated by the dissipative interactions, the
kinetic contribution to the thermal conductivity is subdominant and has the
functional form
\begin{equation}
{\cal L}_{KI} \sim \rho \phi \overline{v} l_{q} \sim \frac{\phi^2 k
T^3}{m L_0 r_{\lambda}^3}    \label{45f}
\end{equation}
Notice that ${\cal L}_{KI}$ scales as $1/L_0$, according to our initial
assumption. 
A similar reasoning leads to the $1/\zeta_0$-dependence of the kinetic
contribution to the viscosity in the region where the momentum transport
is dominated by the friction between particles, as has been already
obtained by different authors\cite{Ern,Gro}. 

        If $L_0$ is reduced, we will eventually have that $l_q > l_p$.
Clearly, in this region the energy transport is dominated by the kinetic
contribution, which is much more effective than the direct transmission of
heat between the particles. It is crucial to realize, however, that 
the flow of particles is dominated by its diffusive motion instead of the
inertial motion of the previous domain, since the momentum has had time to
relax. The characteristic relaxation time
is still given by eq. (\ref{45c}), since the mechanism for energy
delivery keeps on being the direct heat transport between particles.
However, $l_q$ is calculated from the mean displacement of a brownian
particle, that is 
\begin{equation}
l_q \sim \sqrt{Dt_q} \sim \sqrt{\frac{kT}{\zeta_0 \rho r_{\zeta}^3}}
        \label{46a} 
\end{equation}
where the diffusion coefficient has been estimated from a Stokes-Einstein
relation $D \sim kT/\zeta_0 \rho r_{\zeta}^3$. Thus, the characteristic
velocity of this diffusive flow is $l_q/t_q$, from which we obtain the
energy flow in this regime
\begin{equation}
J_q \sim \phi \rho \frac{l_q}{t_q} l_q \nabla T \sim \phi \rho D \nabla T 
        \label{46b}
\end{equation}
Hence, from this expression we infer the kinetic contribution to thermal
conductivity which, in addition, will be the leading contribution in this
region
\begin{equation}
{\cal L}_{KII} \sim \phi \rho D \sim \frac{\phi k T}{\zeta_0
r_{\zeta}^3}    \label{46c}
\end{equation}
The preceeding reasoning leads to a resulting thermal conductivity
independent of the parameter $L_0$ in the region where the dissipative
contribution to the thermal conductivity is small. In this respect, this
new result is
essentially different from that reported for the viscosity coefficient in
the isothermal DPD model\cite{Ern,Pago}. 

\section{Simulation results}

\setcounter{equation}{0}

        The particular system used to investigate the transport properties
of the 
DPDE model consists of $N=10000$ particles in a cubic box of periodic
boundary conditions in all three dimensions, so that no walls
are considered. The lateral size of the box has been set to $N^{1/3}$ in
dimensionless units, and hence the number density $\rho^*$ is always
unity. We have studied essentially dense systems, in the sense that
many particles interact with each other at one time, and therefore, the
interaction ranges are larger than $1/\rho^{1/3}$. The choice of $\rho^* =
1$ also implies that the momentum penetration depth is smaller than, or of
the order of, the mean distance between particles.

Before taking any measurment, the system is allowed to reach 
thermal equilibrium. According to the expressions given in eqs.
(\ref{30}) and (\ref{35e}), and the form chosen in eqs. (\ref{14c}) and
(\ref{14d}) for the dissipative coefficients, the predicted values of the
thermal conductivity and the shear viscosity are, respectively
\begin{eqnarray}
{\cal L} &=&  \frac{2 \pi}{315}\frac{\rho^2}{T^2} L_0 r_{\lambda}^5
\label{40}      \\
\eta &=& \frac{2 \pi}{1575} \rho^2 r_{\zeta}^5 \zeta_0  \label{41}
\end{eqnarray}
Notice the resulting temperature dependence on the thermal conductivity,
which arises from the dependence of the mesoscopic thermal conductivity on the
particles' temperatures.

        The measurement of the transport properties in the system has been
carried out by tracking the relaxation of perturbations externally
induced in the 
system. The hydrodynamic behavior predicted for the DPDE system in the
preceeding section suggests that small perturbations should relax
according to linearised versions of eqs.(\ref{24}), (\ref{28}) and
(\ref{38}). In particular, one finds that a given Fourier component of
wavevector $\vec{k}$ of a
temperature perturbation field in the system should relax as
\begin{equation}
\delta T_k(t) = \delta T_k(0) e^{-\kappa k^2 t}        \label{41c}
\end{equation}
where $\kappa = {\cal L}/c_v \rho$ is the thermal diffusivity, with
\begin{equation}
c_v = \left. \frac{\partial \varepsilon}{\partial T} \right)_v =
\phi\left(1+\frac{5k}{2\phi}\right)      \label{41b} 
\end{equation}
where $c_v$ is the constant volume heat capacity per
particle (related to the value given in eq. (3.22) of ref.\cite{ours2}
according to $c_v=C_v/N$). Notice that $c_v \neq \phi$, due to the
contribution of the energy stored in other degrees of freedom different
from the internal energy of the particle.
Superposed to this relaxation there is also the effect of the temperature
fluctuations induced by the random forces and random heats. However, for 
small values of $B$, the amplitude of the random temperature fluctuations
is very small and, thus, can be ignored. 

        We have obtained the relaxation of temperature fluctuations, $c_v \rho
\delta T_k$, as the difference between perturbations in the internal energy
density $\delta (\rho \varepsilon)_k$, externally induced in the system at
$t=0$, and $c_v T \delta \rho_k $, which 
are the directly observable variables. To obtain the value of the thermal
conductivity, we have 
fitted an exponential decay to the curves obtained
in each simulation. For all the simulations dominated by the dissipative
contribution, we get clear exponential decays, with regression
coefficients typically of $0.999$ or higher, and error bars for the
measured exponent of about 1\% of its value. In the region dominated by
the kinetic contribution, slight deviations from a purely exponential
decay are observed. However, correlation
coefficients typically of $0.98$ and $0.99$ are still found, and the
error is about 10\%. The measured values for the
thermal conductivity as given by means of the procedure described above
have been 
compared with those obtained from the application of a temperature
gradient to the system and measuring the resulting heat flow through the
system once the steady state was reached. The results obtained by both
methods are in agreement.

        We have performed two series of measuremets of the thermal
conductivity for two different radii $r_{\lambda}$. The interaction number
is then given by
\begin{equation}
n \equiv \frac{4 \pi}{3} \rho r_{\lambda}^{3} = \frac{4 \pi}{3} \rho^*
r_{\lambda}^{* 3}     \label{42} 
\end{equation}
We have plotted the results
for ${\cal L}^* \equiv {\cal L} ml/\phi\zeta_0$ against the dimensionless
parameter $X \equiv C r_{\lambda}^{* 5} \rho^{* 2}/T^{* 2}$, for values of
$X$ ranging from $1$ to $1000$, thus covering three decades of this
parameter. The results are then compared with the theoretical prediction,
given in eq. (\ref{40}), that reads
\begin{equation}
{\cal L}^* =\frac{m l}{\phi \zeta_0}  {\cal L} = \frac{2\pi}{315} C
 \frac{\rho^{* 2}}{T^{* 2}} r_{\lambda}^{* 5}    \label{43}
\end{equation}
in dimensionless form. The dimensionless temperature was about $1.1$ and
the density was unity. The range of the interaction $r_{\lambda}^*$
was set to $1.24$ and $2.88$, so that the interaction number was
$n = 8$ and $n=100$, respectively. The dimensionless shear viscosity
coefficient takes the form
\begin{equation}
\eta^* = \eta \frac{l}{\zeta_0}  = \frac{2 \pi}{1575}r_{\zeta}^{* 5}
\rho^{* 2} \label{44} 
\end{equation}
According to our set of dimensionless variables, the viscosity coefficient
depends only on the density and interaction range and, hence, is constant
for each set of simulations characterized by a given value of $n$. We have
measured a viscosity coefficient $\eta^* = 0.363$ for the case $n=8$ and
$\eta^* =0.894$ for $n=100$. We set the parameter $B$ to $10^{-5}$ to avoid
negative values of the energy during the
simulation for 
large $C$ values. Likewise, the time-step was chosen to be $\delta t^* =
10^{-2}$ 
or $\delta t^* = 10^{-3}$ in the critical cases regarding the
possibility of negative values of the particles' energy.

        In fig. 1 we plot our results for the thermal conductivity as a
function of the parameter $X$. For both values of $n$ we observe
a linear 
dependence of ${\cal L}^*$ in this parameter, for values of $X$ larger than
$315/2\pi \simeq 50$, which roughly corresponds to the point at which the
thermal conductivity is about $1$. Thus, we can conclude that the
dissipative contribution dominates the thermal conductivity in this
region, and that the dependence in $X$ is well captured by eq. (\ref{43}).
In addition, for the case $n =
100$, we find a much better agreement between the theoretical predictions
and the simulation results than in the case $n = 8$, whose value is
roughly half of the theoretical one in all of the range. This is a rather
unexpected result which is caused neither by the particle number fluctuations
in the sample nor by significant deviations of $g(r)$ from $1$, as we have
checked. Such a
behavior is equally independent of the way of measuring the property,
since good agreement has been found between the data obtained from both
the relaxation method and heat flow in a temperature gradient.
Furthermore, data reported in ref.\cite{Esp3} for a system of frozen DPDE
particles also show the same kind of dependence of the thermal
conductivity with $n$, so that it cannot be attributed to a possible
coupling between energy and momentum transport. This effect has no
equivalent in the case of the viscosity coefficient which, in our
dimensionless variables, varies only with the range of the interaction
$r_{\zeta}^*$ at constant $\rho^*$. 

        In fig. 2 we focus our attention on the range $0 \leq X \leq 50$,
so that the dissipative contribution to the thermal conductivity is
subdominant. We plot the results for the two series of data ($n=8$ and
$n=100$) as a function of
the parameter $Y \equiv T^{* 3}/Cr_{\lambda}^{* 3} \sim {\cal L}_{K
I}^*$, to compare with the prediction given in eq. (\ref{45f}). Both series
of data agree with a thermal conductivity independent of $L_0$ ($C$ in
dimensionless variables) in this range of values of $L_0$. Although the
data are less reliable than in the previous range, we can consider that
the behavior given in eq. (\ref{46c}) is confirmed and that the transport of
energy for small $L_0$ is due to particle diffusion.
We should point out, however, that the data
with $n=8$ show an increase of the thermal conductivity in passing from
the dissipative regime to the diffusive regime, while the data for $n=100$
shows a smooth crossover. This can attributed to the fact that for $n=8$ the
crossover between both regimes passes first through the behavior described
in eq. (\ref{45f}). In the $n=100$ case, the dissipative and
diffusive regimes simply overlap in the crossover. 

        To end this section, we report data on the shear viscosity
coefficient obtained for the model. According to our choice of parameters,
the shear viscosity is only a function of the range of interaction
$r_{\zeta}$. Thus, we have performed simulations and plotted in fig. 3 the
viscosity coefficient as a function of the parameter $Z \equiv \rho^{* 2}
r_{\zeta}^{* 5}$. As expected, we obtain a good agreement between
the theoretical prediction given in eq. (\ref{44}) for values of the
parameter $Z$ larger than $1575/2\pi \sim 250$, which roughly corresponds
to a viscosity $\eta^* \simeq 1$. In this region, however, a slight
curvature can be observed. For values of $Z$ smaller than $250$, as is to
be expected, the momentum transport is dominated by kinetic effects. Here, we
observe the same kind of qualitative behavior as found for the isothermal
DPD model as has been discussed elsewhere\cite{Ern,Pago,War}.

\section{Conclusions}

\setcounter{equation}{0}

        In this paper we have analyzed the transport properties of the
new DPDE model from both, a simple theoretical approach and computer
simulations. 
This model is addressed at the simulation of fluctuating fluids in which
momentum as well as heat transport are involved, and has already shown well
defined equilibrium properties\cite{ours2}. In this respect, the
model is {\em complete} in the sense that it is thermodynamically consistent
and the five hydrodynamic fields can
be correctly described. The complete understanding of its features and the
refinement of its capabilities, however, still demands a great deal of
additional effort.

        After a brief review of the algorithm, we have first of all
derived the dynamic properties of the DPDE model 
based on a local equilibrium assumption. In this framework, we have
obtained the transport equations and 
approximate expressions for the transport coefficients which here
can be functions of the temperature. In this approach, it is
implicitly assumed that an
initial probability distribution for the relevant variables of the system
will relax to the local equilibrium form after a time $t_p$ for the
momentum and $t_q$ for the energy, according to eqs. (\ref{45}) and
(\ref{45c}), respectively. These characteristic times must be much smaller
than those related to the changes in the hydrodynamic fields for the whole
approach to be valid. We have qualitatively argued that the local
equilibrium is the dominant contribution in an expansion in inverse powers
of the dissipative coefficients $L_0$ and $\zeta_0$, so that 
the transport coefficients
can be expressed as series expansions of this particular form. This is
also the case in the derivation of 
the Smoluchowski equation from the Fokker-Planck
equation\cite{vK,JB4,JB5}. Therefore,
in the range of validity of this point of view, the kinetic contributions are
always subdominant. The analytic form for these subdominant kinetic
contributions has been 
reported in the literature\cite{Ern,Gro} in the case of the viscosity, and
are given in eq. (\ref{45f}) in the case of the thermal conductivity.

        Second, we have also studied the behavior of the system when the
dissipative contribution becomes comparable or even smaller than the
kinetic contribution to the dissipative coefficients. In this regime, the
theoretical picture based on the local equilibrium assumption fails. It is
important to realize, however, that the form of the series expansion
of the transport coefficients in inverse powers of $L_0$ and $\zeta_0$
will change towards another functional dependence. This is in fact one of
the more 
important findings of this paper which also applies in the case of
viscosity in the
isothermal DPD model. In the case of the thermal conductivity
we have found that the thermal conductivity at small values of $L_0$
should be driven by particle diffusion, according to eq. (\ref{46c}). We find
that ${\cal L}$ is independent of $L_0$ and that the simulations performed
seem to agree with this interpretation. Thus, the increase of the thermal
conductivity at small values of $L_0$ is attributed to a crossover from
the regime dominated by the dissipative interactions to a regime dominated by
diffusive transport. This crossover region depends on the interaction
number $n$.

        Third, we have shown the first simulation results for both, the shear
viscosity coefficient and the thermal conductivity for the DPDE model.
Qualitatively, the 
shear viscosity for the DPDE behaves in the same way as that of the
isothermal DPD 
model, as could have been expected. Furthermore, we have explored
different parameter regions that cover only part of all the
possibilities of the DPDE model. Particular attention
has been paid to non-dilute 
systems, aiming at a reliable comparison between the simulation data and
the theoretical results in 
the large friction limit. We have also reported results in regions where the
theoretical assumptions fail and new qualitative behaviors have been
found for the thermal conductivity. 

        Fourth and last, an important result of our simulation analysis is
the difference (of about a factor of two between the two series of data) 
observed in the thermal conductivity, in the region of large $L_0$, for
the two values of the interaction number analyzed. We find no qualitative
explanation for this difference, which becomes larger as the number of
particles inside the interaction range diminishes. The origin of this
discrepancy could be the same as that found for the viscosity in
the isothermal DPD model, according to ref.\cite{War}. The fact that the
CPU time grows with the interaction 
number, makes the regime of small $n$ be of great interest by itself. 

\section{Acknowledgements}

        This work has been partially supported by the Direcci\'on General
de Ciencia y Tecnolog\'{\i}a of the Spanish Government under the contract
PB96-1025, as well as funding from the Generalitat de Catalunya (ACES
1999). The authors would like to thank V. Navas and M.C. Molina for their
interest in this work.

\newpage

\section*{Figure Captions}

\noindent Fig. 1: Dimensionless thermal conductivity as a function of
$X = Cr_{\lambda}^{* 5} \rho^{* 2}/T^{* 2}$. Squares stand for data for
$n=8$ while circles represent data for $n=100$. The solid line is the
theoretical prediction given in eq.(\ref{43})

\noindent Fig. 2: Dimensionless thermal conductivity as a function of
$Y = T^{* 3}/Cr_{\lambda}^{* 3}$. Squares stand for data for
$n=8$ while circles represent data for $n=100$. The inset shows a detailed
view of the $n=8$ data.

\noindent Fig.3: Dimensionless shear viscosity as a function of $Z=
\rho^{* 2} r_{\zeta}^{* 5}$. The solid line stands for the theoretical
prediction given in eq. (\ref{44}), while the circles indicate simulation
data points.


\begin{thebibliography}{99}
\bibitem{vK3} The It\^o prescription of the non-linear Langevin equations
would lead to an algorithm similar to that of eqs.(\ref{21a}), (\ref{21b})
and 
(\ref{21c}) where the primed functions are calculated at the time $t$.
Instead, Stratonovich would prescribe the calculation of the same
functions at the time $t+\delta t/2$. See ref.\cite{vK}.
\end{thebibliography}
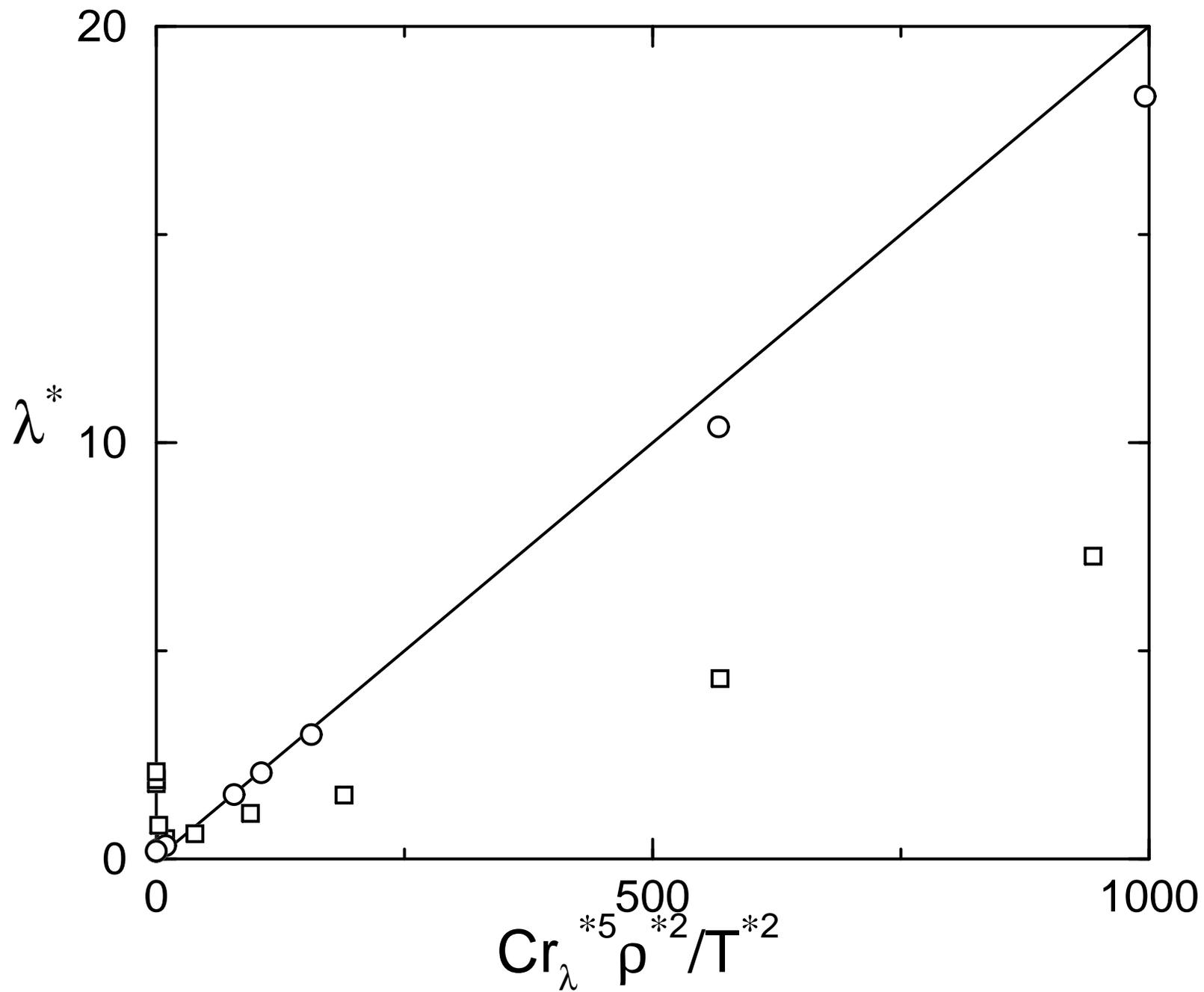
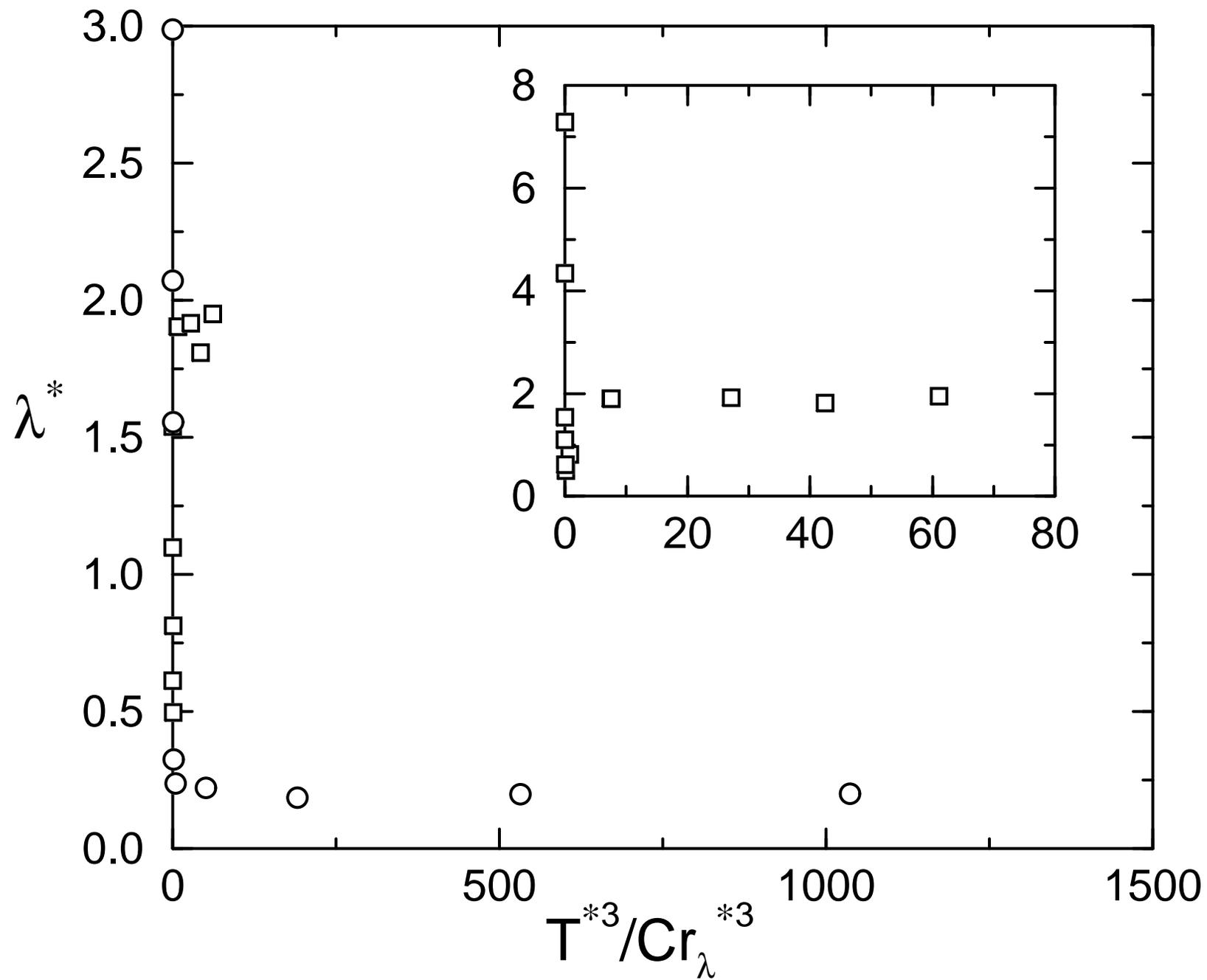
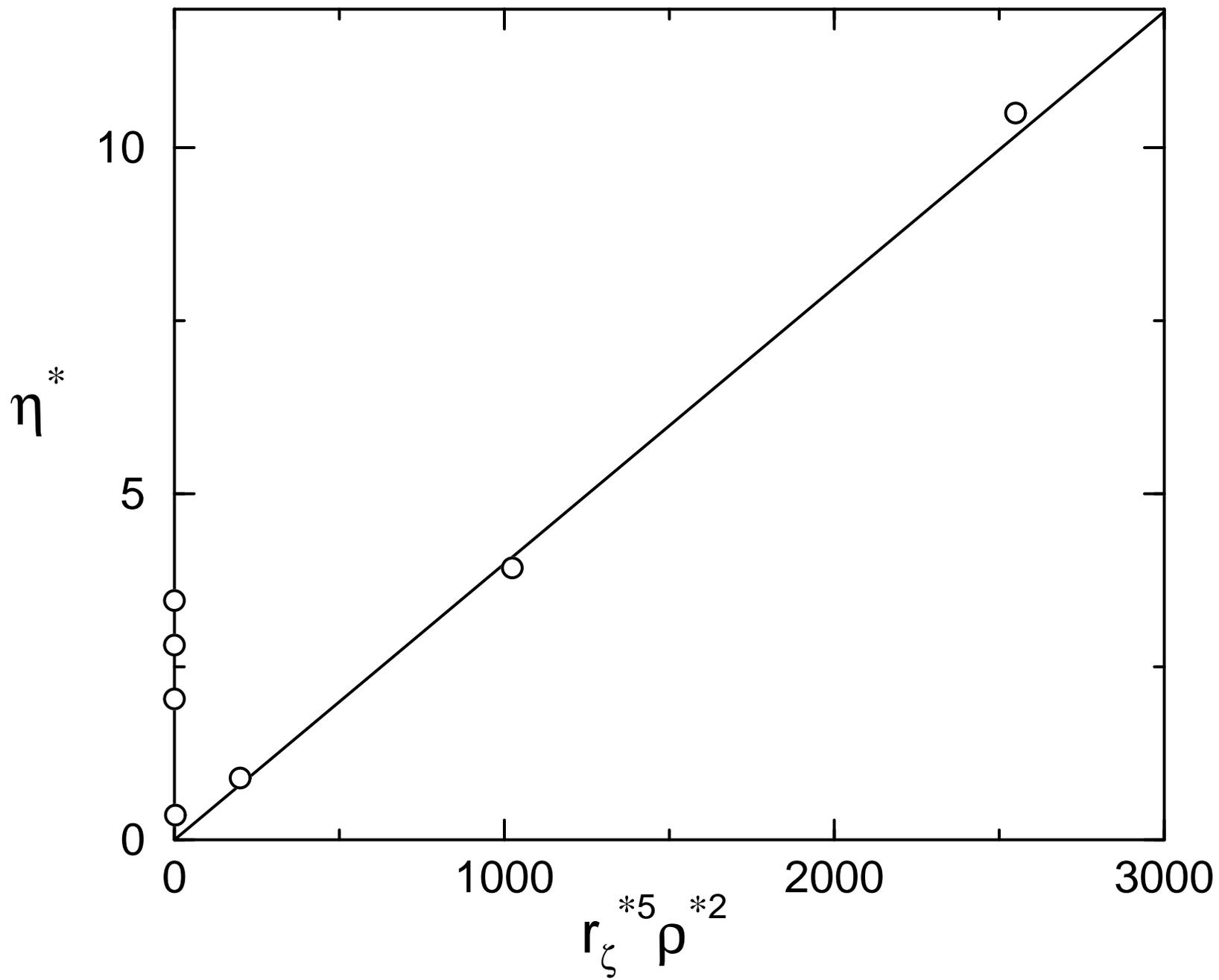
\end{document}